\documentclass[superscriptaddress,prl,amsmath,amssymb,eufrak,aps,floatfix,twocolumn,notitlepage,longbibliography]{revtex4-2}

\usepackage{lipsum}
\usepackage{braket}
\usepackage{graphicx}
\usepackage{dcolumn}
\usepackage{enumitem}
\usepackage{bm}
\usepackage{amsmath}
\usepackage{amssymb}
\usepackage{mathtools}
\usepackage{subfigure}
\usepackage{color}
\usepackage{xcolor}
\usepackage{verbatim}
\usepackage{float}
\usepackage{appendix}
\usepackage{bbold}

\newcommand{\abs}[1]{\lvert #1 \rvert} 

\usepackage{upgreek}
\usepackage{physics}
\usepackage{isomath}
\usepackage{placeins}
\usepackage{hhline}
\usepackage{hyperref}
\usepackage{cancel}

\begin{document}

\preprint{APS/123-QED}

\title{Collective Excitations of Dissipative Time Crystals}

\author{Gage W. Harmon}
\affiliation{Theoretische Physik, Saarland University, Campus E2.6, 66123 Saarbrücken, Germany}
\author{Giovanna Morigi}
\affiliation{Theoretische Physik, Saarland University, Campus E2.6, 66123 Saarbrücken, Germany}
\author{Simon B. J\"ager}
\affiliation{Physikalisches Institut, University of Bonn, Nussallee 12, 53115 Bonn, Germany}
\date{\today}

\date{\today}

\begin{abstract}
	We study the dynamics of atoms interacting periodically with a dissipative optical cavity and employ Floquet theory to analyze their low-frequency behavior. By means of an effective atom-only master equation, valid in the bad cavity regime, we characterize the excitation spectrum of the atoms across the transition from a normal phase to a time-crystalline phase where the atoms undergo stable oscillations. We identify features in the complex excitation spectra when crossing second and first order transitions where the order parameter changes continuously or abruptly. Finally, we discuss how these results can be detected experimentally by probing the cavity with an additional drive. Our work provides important tools for analyzing the response of dynamical out-of-equilibrium phases.  
	
\end{abstract}

\maketitle
\emph{Introduction---} 
Driven atoms coupled to a dissipative optical cavity provide a rich platform for exploring physics beyond equilibrium and ground-state descriptions. Examples of observed out-of-equilibrium phenomena include, among others: self-organization \cite{Domokos:2002,Nagy2008,Baumann:2010,Arnold:2012,Ritsch:2013,Schuetz:2015,Kroeze:2018,Landig:2016,Halati:2020,Helson:2023,Wu:2023}; synchronization \cite{Xu:2014,Weiner:2017,Nadolny:2025,Natale:2025}; self-oscillating pumps \cite{Lin:2022,Rosa-Medina:2022,Dreon:2022,Zhang:2025}; cold-atom lasing \cite{Guerin:2008,Bohnet:2012,Gothe:2019,Kristensen:2023,Schäfer:2025}; and both continuous and discrete dissipative time crystals \cite{Piazza:2015,Iemini:2018,Chitra:2015,Gong:2018,Kessler:2021,Kongkhambut:2022,Kongkhambut:2024}. The latter two are remarkable examples of out-of-equilibrium phases of matter that can exist only because of the beneficial interplay between driving and dissipation. Although many works have focused on the emergence and description of the corresponding quantum states, very little is known about their low-frequency excitations and spectral properties. Closing this gap will not only provide insight into the universality of these phases but also enable their use in quantum technologies.  

A widely studied class of time crystals are discrete dissipative time crystals (DTCs): periodically driven systems that spontaneously break the time-translational symmetry imposed by the drive and stabilized by dissipation. The existence of periodic driving 
circumvents the no-go-theorem for time crystals~\cite{Wilczek:2012,Sacha:2018,Bruno:2013,Watanabe:2015} and DTCs, in particular, can be studied in the dissipative Dicke model. This model describes strong light-matter interactions where the coupling is dynamically modulated to parametrically excite the underlying states of light and matter \cite{Chitra:2015,Gong:2018,Kessler:2021}. Remarkably, this resonant excitation does not heat the system on long timescales, because dissipation stabilizes and guides it into a limit cycle. This phase has been realized experimentally~\cite{Kessler:2021}, where it was shown to be robust against perturbations. More recently, works have explored such systems with a focus on topological invariants~\cite{Villa:2024,Wahl:2024}, exotic bifurcations~\cite{Jaeger:2019,Liu:2025,Cosme:2025}, and quantum metrology~\cite{Montenegro:2023,Mattes:2023,Gribben:2025}. Surprisingly, however, the low-frequency behavior across this phase transition has been widely overlooked, even for the simplest systems. Resolving this issue is important not only for a fundamental understanding of out-of-equilibrium transitions but also, more generally, for providing theoretical tools and experimental protocols to reveal and analyze more involved systems.

In this Letter, we analyze the low-frequency excitations of DTCs across transitions. This approach is complementary to previous approaches that use bifurcation theory to classify out-of-equilibrium transitions~\cite{Kosior:2023,Liu:2025,Skulte:2024,Cosme:2025}. Bifurcation theory provides a useful tool to explore mode softening~\cite{Nie:2023} from one side of the transition towards an instability across the transition. In contrast, we employ self-consistent Floquet theory to extract the complex excitation spectrum of the stable phases across transitions. Our approach follows the logic of classifying transitions by their Liouvillian gap~\cite{Minganti:2018}, but for periodically-driven systems where exact diagonalization of the Liouvillian is computationally expensive~\cite{Harmon:2022}. We examine (i) continuous transitions and (ii) discontinuous transitions, in which the time-averaged order parameter changes smoothly or exhibits a jump, respectively. We further show how these excitations can be probed experimentally by using truncated Wigner simulations to model the measurement process. From the dynamical response we extract the frequencies and linewidths and compare them with the Floquet theory predictions. Our work provides theoretical tools for analyzing out-of-equilibrium phase transitions. 

\begin{figure}[!tb]
	\centering
	\includegraphics[width=1\columnwidth]{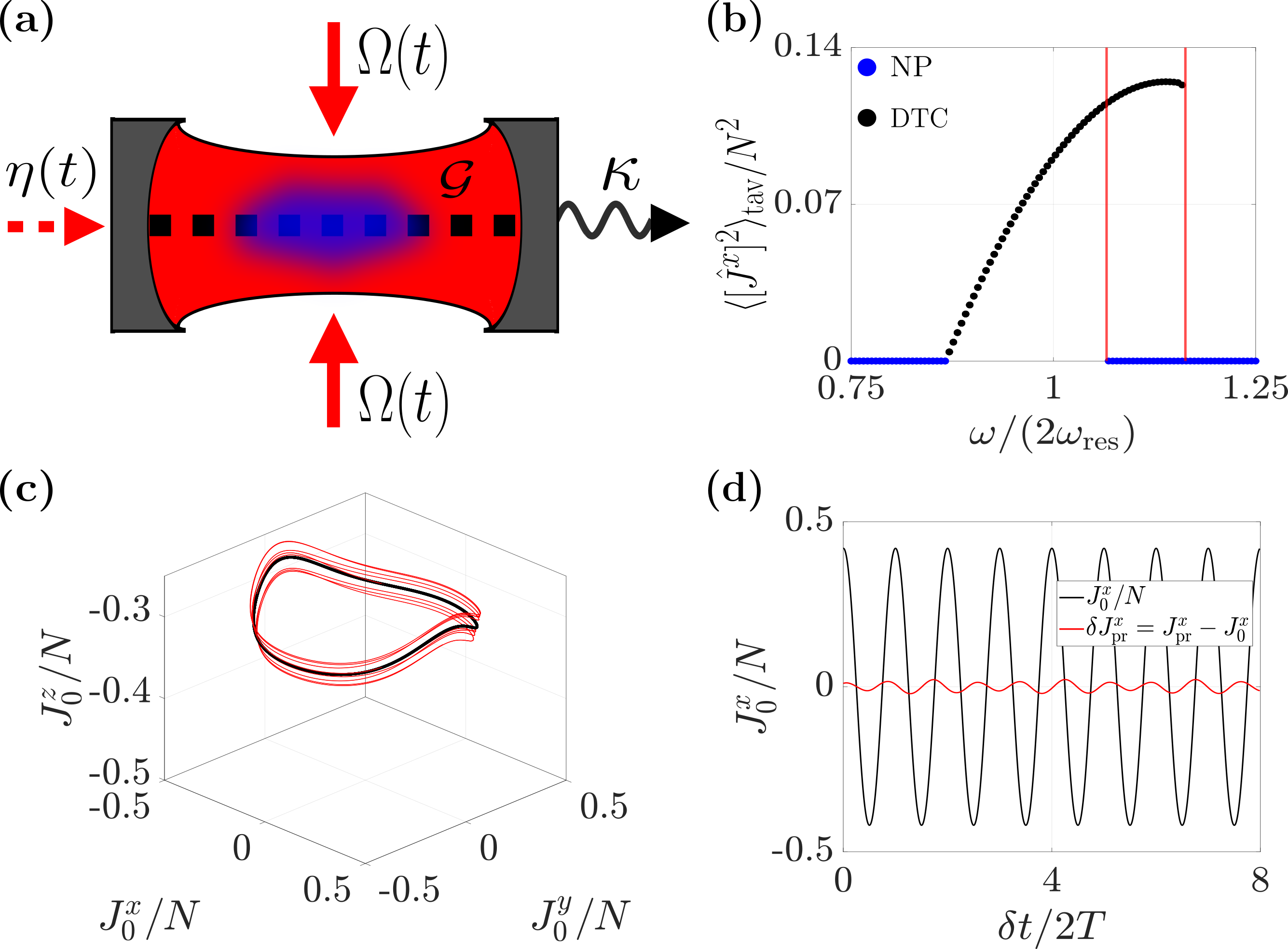}
	\caption{(a) Schematic of an experimentally realizable setup where a cloud of ultracold atoms are driven by two red-detuned lasers with modulated Rabi frequencies and are coupled to a single dissipative mode. An additional drive $\eta(t) = \eta_0e^{-i\omega_\text{pr}t}$ weakly probes the atoms in order to benchmark the results found in Fig.~\ref{fig:fig2} and can be shown in Fig.~\ref{fig:fig3}. (b) Order parameter as a function of driving frequency. (c) Mean-field trajectory of $J^x_0(t), J^y(t), J^z_0(t)$ for $0 \leq \delta t/2T \leq 8$ calculated from Eqs.~\eqref{eq3}-\eqref{eq5} with probe strengths $\eta_0 = 0\kappa$ (black) and $\eta_0 = 0.1\kappa$ (red). (d) Mean-field trajectory of $J^x_0(t)$ with $\eta_0 = 0\kappa$ (black) and $\delta J^x_\mathrm{pr}(t) = J^x_\text{pr}(t) - J^x_0(t)$ with $\eta_0 = 0.1\kappa$ (red). The remaining parameters are $g_1/g_0 = 0.6, \, \delta_c = \kappa, \, \omega = 2\omega_\mathrm{res}, \, \omega_\mathrm{pr} = \omega/2 + 0.19\omega_\mathrm{res}, \, g_0/g_c = 0.5,$ and $\Delta = 0.1\kappa$.}
	\label{fig:fig1}
\end{figure}
\emph{Dynamical description.---} 
We consider a transversely driven gas of $N$ atoms with two ground states, $\ket{\downarrow}$ and $\ket{\uparrow}$, that couple strongly to an optical cavity [Fig.~\ref{fig:fig1}(a)]. The laser and cavity couple the two ground states to an excited-state manifold, which can be eliminated by driving far off resonance. The effective model is captured by the master equation that governs the evolution of the density operator, $\hat{\rho}$, of the cavity and ground-state degrees of freedom ($\hbar = 1$)
\begin{equation}
	\begin{aligned}\label{eq1}
		\frac{\partial \hat{\rho}}{\partial t} = -i\big[\hat{H},\hat{\rho}\big] - \kappa(\hat{a}^\dagger\hat{a}\hat{\rho} + \hat{\rho}\hat{a}^\dagger\hat{a} - 2\hat{a}\hat{\rho}\hat{a}^\dagger).
	\end{aligned}
\end{equation}
The Hamiltonian is given by 
\begin{equation}
	\begin{aligned}\label{eq2}
		\hat{H} = \delta_c\hat{a}^\dagger\hat{a} + \Delta\hat{J}^z + \frac{g(t)}{\sqrt{N}}\big(\hat{a} + \hat{a}^\dagger\big)\big(\hat{J}^+ + \hat{J}^-\big).
	\end{aligned}
\end{equation}
Here, we introduce the cavity annihilation (creation) operators, $\hat{a}$ ($\hat{a}^\dagger$); the collective spin-lowering (raising) operators, $\hat{J}^- = \sum_{k=1}^{N} \hat{\sigma}^-_k$ (with $[\hat{J}^-]^\dagger = \hat{J}^+$); and the collective spin operators, $\hat{J}^a = \sum_{j=1}^{N} \hat{\sigma}^a_j / 2$ ($a = x, y, z$), with
$\hat{\sigma}^-_j = \ket{\downarrow}_j \!\bra{\uparrow}_j$,
$\hat{\sigma}^x_j = (\hat{\sigma}^+_j + \hat{\sigma}^-_j)/2$,
$\hat{\sigma}^y_j = i(\hat{\sigma}^-_j - \hat{\sigma}^+_j)/2$, and
$\hat{\sigma}^z_j = (\hat{\sigma}^+_j \hat{\sigma}^-_j - \hat{\sigma}^-_j \hat{\sigma}^+_j)/2$.
Furthermore, $\kappa$ is the cavity linewidth; $\delta_c = \omega_c - \omega_l$ is the detuning between the cavity and the laser driving frequency; and $\Delta$ is the frequency gap between the ground states, which can be internal~\cite{Dimer:2007,Baden:2014} or external~\cite{Baumann:2010,Kessler:2021} to the atoms. The coupling between the atoms and the cavity $g(t) = g_0 + g_1 \cos(\omega t)$ is time-periodic. In experimental settings, this is realized by detuning frequency tones in the driving amplitude $\Omega(t)$ by an amount $\omega$ from the carrier frequency $\omega_l$. As a result of this, the cavity-enhanced Raman scattering rate $\propto g(t)$ between the ground states becomes modulated in time~\cite{Kessler:2021}. 

The core idea of the DTC transition is to tune the modulation frequency $\omega$ to the parametric resonance $\omega \approx 2\omega_{\mathrm{res}}$ of the lower-polariton mode
\begin{align}
	\omega_{\mathrm{res}}=\Delta\sqrt{1-\frac{g_0^2}{g_c^2}},
\end{align}  
with $g_c = [\Delta(\delta_c^{2} + \kappa^{2})/(4\delta_c)]^{1/2}$. This allows for a resonant pair-creation process that results in a subharmonic response of the Raman-scattered light field. Remarkably, even on resonance, this emission stabilizes due to dissipation, and the system reaches a superradiant state that exhibits stable subharmonic oscillations. The transition can be seen in the time-averaged order parameter $\langle[\hat{J}^{x}]^{2}\rangle_{\mathrm{tav}} = \int_{0}^{T}\!dt\,\langle[\hat{J}^{x}(t)]^{2}\rangle/T$, plotted in Fig.~\ref{fig:fig1}(b) as a function of the driving frequency $\omega$. When $\omega$ is sufficiently close to the parametric resonance, we find a transition from the normal, non-superradiant phase (NP) to a superradiant DTC that spontaneously breaks the discrete $\mathbb{Z}_{2}$ symmetry [$\,(\hat{J}^{x},\hat{a}) \mapsto -(\hat{J}^{x},\hat{a})$] and the discrete time-translational symmetry ($\,t \mapsto t + T$, $T = 2\pi/\omega$). We find two kinds of transitions: (i) a continuous transition, where the order parameter varies smoothly for $\omega < 2\omega_{\mathrm{res}}$, and (ii) a discontinuous transition, where the order parameter jumps and we find bistable, non-superradiant, and superradiant solutions for $\omega > 2\omega_{\mathrm{res}}$. Within the superradiant phase, the atomic state exhibits a limit cycle, visible in Fig.~\ref{fig:fig1}(c) as a solid black curve, and the non-time-averaged order parameter (also in black) exhibits stable subharmonic oscillations, as shown in Fig.~\ref{fig:fig1}(d). The main objective of this Letter is to quantitatively describe excitations around these limit cycles and their dynamics across the different transitions. Such excitations have finite lifetimes and resonance frequencies that are incommensurate with the many-body oscillating DTC rotating at $\omega/2$. An example of this is displayed in red in Figs.~\ref{fig:fig1}(c) and (d). 

Since we are interested in the dynamics of the atomic excitations, we employ an atom-only description of the DTC~\cite{Jaeger:2022,Jaeger:2024,Harmon:2025}. This description relies on a timescale separation between the cavity and atomic degrees of freedom. In particular, we assume that the typical cavity timescale $\tau_c \sim |\delta_c + i\kappa|^{-1}$ is much shorter than the typical atomic timescale determined by the lower-polariton resonance $\tau_a \sim |\omega_{\mathrm{res}}|^{-1}$, that is $\tau_c \ll \tau_a$. Consistent with this limit, we work in the regime where the driving frequency $\omega$ cannot resonantly excite photonic-type excitations $\omega \tau_c \ll 1$, but can excite matter-type excitations $\omega \tau_a \sim 1$. Following the approach of Refs.~\cite{Jaeger:2024}, we derive an atom-only Lindblad master equation that encapsulates the cavity-mediated, time-periodic interactions and dissipation. Additionally, we perform a mean-field decoupling by replacing the operators $\hat{J}^{a}$ with their expectation value, $J^{a}$ ($a = x, y, z$). We can then describe the dynamics of the mean-field spin variables with the following non-linear equations
\begin{align}
	\frac{dJ^x}{dt} =& -\Delta J^y,\label{eq3}\\
	\frac{dJ^y}{dt} =& \Delta J^x + \frac{4V_0 J^xJ^z}{N} + \frac{4V_1J^yJ^z}{N},\label{eq4}\\
	\frac{dJ^z}{dt} =& -\frac{4V_0J^xJ^y}{N} - \frac{4V_1[J^y]^2}{N}.\label{eq5}
\end{align}
The detailed derivation is shifted to the Supplemental Material (SM)~\cite{SM}. Here, the terms proportional to $V_0 = 2g^{2}(t)\delta_c/(\delta_c^{2}+\kappa^{2}) - 4g(t)\dot{g}(t)\delta_c\kappa/(\delta_c^{2}+\kappa^{2})^{2}$ describe the coherent, cavity-mediated interactions, and the terms proportional to $V_1 = 4g^{2}(t)\delta_c\kappa/(\delta_c^{2}+\kappa^{2})^{2}$ arise from cavity-mediated dissipation. In a previous Letter, it was shown that both terms are required to describe the DTC and the stabilization of the limit cycles~\cite{Jaeger:2024}. The results shown in Figs.~\ref{fig:fig1}(b) and \ref{fig:fig1}(d) are obtained from these equations. 

\begin{figure}[!tb]
	\centering
	\includegraphics[width=0.45\textwidth]{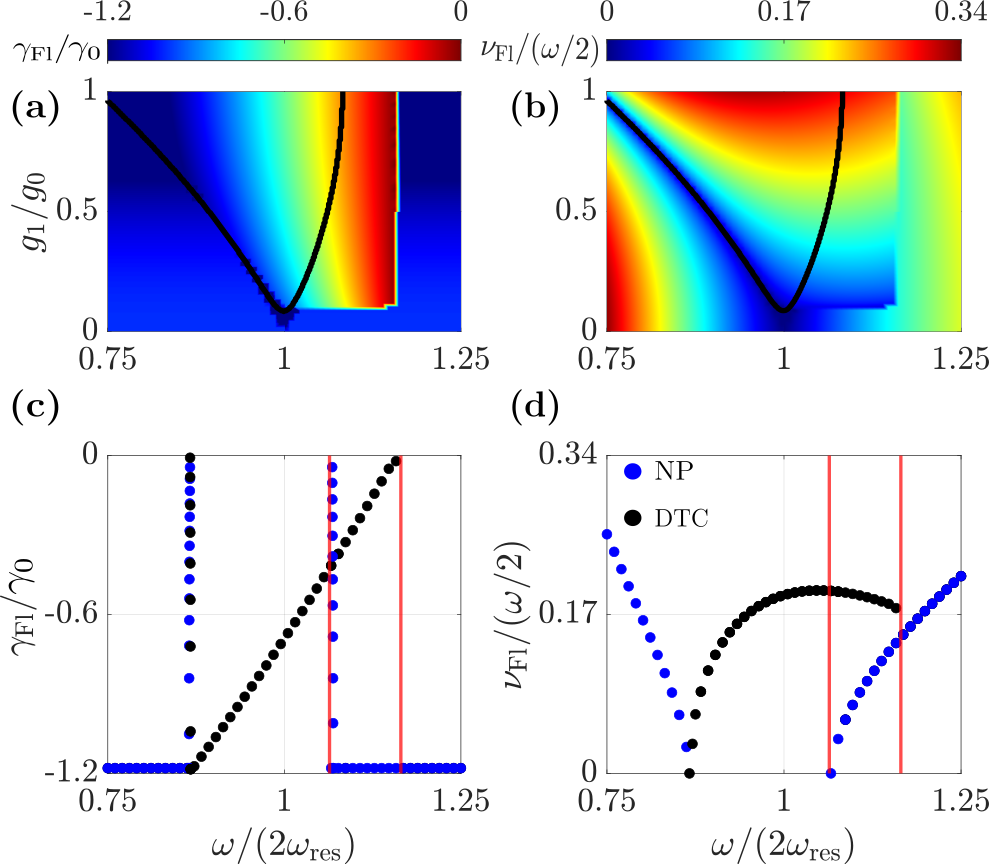}
	\caption{(a) Damping rate $\gamma_\mathrm{Fl}$ in units of $\gamma_0$ and (b) oscillation frequency $ \nu_\mathrm{Fl}$ in units of $(\omega/2)$, both plotted as a function of the driving frequency $\omega/(2\omega_\mathrm{res})$ and modulation strength $g_1/g_0$. (c) Horizontal cut of (a) at $g_1/g_0 = 0.6$. (d) Horizontal cut of (b) at $g_1/g_0 = 0.6$. The black contour in (a) and (b) denotes the phase boundary between NP and DTC phase $(\gamma_\text{Fl} = 0)$ . In (c) and (d) the blue dots are in the NP, black dots are in the DTC phase, and the vertical red bars mark the region of bistability. The other parameters are $\delta_c = \kappa, \, g_0/g_c = 0.5,$ and $\Delta = 0.1\kappa$.}
	\label{fig:fig2}
\end{figure}       

\emph{Describing Excitations.---} 
We will now utilize this description to extract information about the excitations around the limit cycles. This is achieved in two steps. First, we simulate Eqs.~\eqref{eq3}, \eqref{eq4}, and \eqref{eq5} for sufficiently long times such that the system has reached either the NP or the DTC phase; we denote this state as $\vec{v}_0 = (J_0^x(t), J_0^y(t), J_0^z(t))$. Second, we study small fluctuations $\delta\vec{v} = (\delta J^x, \delta J^y, \delta J^z)$ around this state by linearizing the equations of motion. Consequently, we obtain a linear theory $\partial_t \delta\vec{v} = \Sigma\,\delta\vec{v}$ describing the fluctuations with
\begin{equation}\label{eq6}
	\begin{aligned}
		\Sigma = \begin{pmatrix}
			0 & -\Delta & 0 \\ \\ \Delta + \frac{4V_0J^z_0}{N} & \frac{4V_1J^z_0}{N} & \frac{4V_0J^x_0}{N} + \frac{4V_1J^y_0}{N} \\ \\ -\frac{4V_0J^y_0}{N} & -\frac{4V_0J^x_0}{N} - \frac{8V_1J^y_0}{N} & 0   
		\end{pmatrix},
	\end{aligned}
\end{equation}
that depends on the solution $\Vec{v}_0$. The matrix $\Sigma$ is time-dependent due to the explicit $T$-periodic time dependence of $V_{0}$ and $V_{1}$, but also due to the mean-field solution, $\Vec{v}_{0}$, which is itself $2T$-periodic. Thus, we can use Floquet theory to capture all excitations described by $\Sigma$. In practice, this is done by calculating
\begin{equation}\label{eq8}
	\begin{aligned}
		\Phi = \mathcal{T} \exp\int_{0}^{2T} d\tau\,\Sigma(\tau),
	\end{aligned}
\end{equation}
where $\mathcal{T}$ is the time-ordering operator. Subsequently, we can extract the Floquet eigenvalues $\lambda_j$ by finding the eigenvalues $\phi_j$ of $\Phi$ and identifying them with $\phi_j = e^{2\lambda_j T}$. Note that there are three Floquet eigenvalues; one of them, $\lambda_{0} = 0$, is a direct consequence of the conserved spin length, $[J_{x}^{0}]^2 +[J_{y}^{0}]^2 + [J_{z}^{0}]^2=$const. . The other two eigenvalues are non-trivial but appear as complex-conjugate pairs, which allows us to classify the excitation by a single complex number $\lambda_{\mathrm{Fl}} = \gamma_{\mathrm{Fl}} - i\nu_{\mathrm{Fl}}$ where $\nu_{\mathrm{Fl}}$ and $\gamma_{\mathrm{Fl}}$ are the excitation frequency and linewidth, respectively. The method for finding $\lambda_{\mathrm{Fl}}$ is efficient and allows us to map out the excitations over a wide range of parameters (see SM \cite{SM} for details). 

In Fig.~\ref{fig:fig2}(a), we plot $\gamma_{\mathrm{Fl}}/\gamma_0$, where $\gamma_0 = 4g_0^{2}\delta_c\kappa\Delta/(\delta_c^{2} + \kappa^{2})^{2}$ is the time-averaged damping rate, as a function of $g_1/g_0$ and $\omega/(2\omega_{\mathrm{res}})$. The phase boundary (black contour) indicates when the NP becomes unstable. The left boundary (i) is the continuous transition, and the right boundary (ii) is the discontinuous transition. Outside the phase boundary, the NP is stable with $J_0^{z} = -N/2$ and $J_0^{x} = 0$. Within the phase boundary, the system reaches a stable DTC phase and oscillates at $\omega/2$. The red-colored parameter region to the right of the phase boundary is the regime of bistability, which means a DTC solution coexists with the NP. In Fig.~\ref{fig:fig2}(b), we plot the excitation frequency in units of $\omega/2$ for the same parameters. A first finding is that the left phase boundary (i) is accompanied by a vanishing frequency, reminiscent of a second-order quantum phase transition. At the right boundary (ii), we observe discontinuous behavior, as expected for a first-order phase transition.

A deeper understanding of the transition can be drawn from cuts of Fig.~\ref{fig:fig2}(a) and (b) that are visible in (c) and (d) for $g_1/g_0 = 0.6$. In Fig.~\ref{fig:fig2}(c), we find that the destabilization of the NP (blue dots) appears in a very narrow parameter region. In that same region, the excitation frequency $\nu_{\mathrm{Fl}}$ is locked at the value $0$. The dynamics of the DTC close to the phase boundaries is visible as the black points. At (i), the continuous phase boundary, we observe very similar behavior where the frequency $\nu_{\mathrm{Fl}}$ vanishes and $\gamma_{\mathrm{Fl}}$ is rapidly changing over a narrow parameter regime. At this transition point, $\nu_{\mathrm{Fl}}$ has the characteristic ``cusp-like'' non-analytic behavior reminiscent of second-order phase transitions. Note, however, that the cusp itself is in a narrow critical parameter regime where $\gamma_{\mathrm{Fl}}$ quickly drops to zero. Increasing $\omega$ from (i), the left phase boundary, we observe a slow destabilization of the DTC phase indicated by a slowly increasing $\gamma_{\mathrm{Fl}}$. We explain this slow destabilization by an increasing value of $\int_{0}^{2T} d\tau\,J^z_0(\tau)/(2T)$, meaning that more atoms remain on average in the excited state $\ket{\uparrow}$. Consequently, the damping, which tries to stabilize $\ket{\downarrow}$, becomes less efficient. This slow increase of $\gamma_{\mathrm{Fl}}$ continues into (ii), the region of bistability [within the red vertical lines in Fig.~\ref{fig:fig2}(c) and (d)], until the DTC becomes critical, $\gamma_{\mathrm{Fl}} = 0$. We highlight that the slope of $\gamma_{\mathrm{Fl}}$ approaches zero completely differently at the left phase boundary compared to the right, while in the DTC phase. In addition, at the discontinuous transition (ii), the values $\nu_{\mathrm{Fl}}$ for the NP and DTC phase do not cross and remain different frequency branches. This comparison shows how the out-of-equilibrium transition we observe both differs from and resembles equilibrium phase transitions, and it stands as one of the main results of this Letter.

\begin{figure}
	\centering
	\includegraphics[width=1\columnwidth]{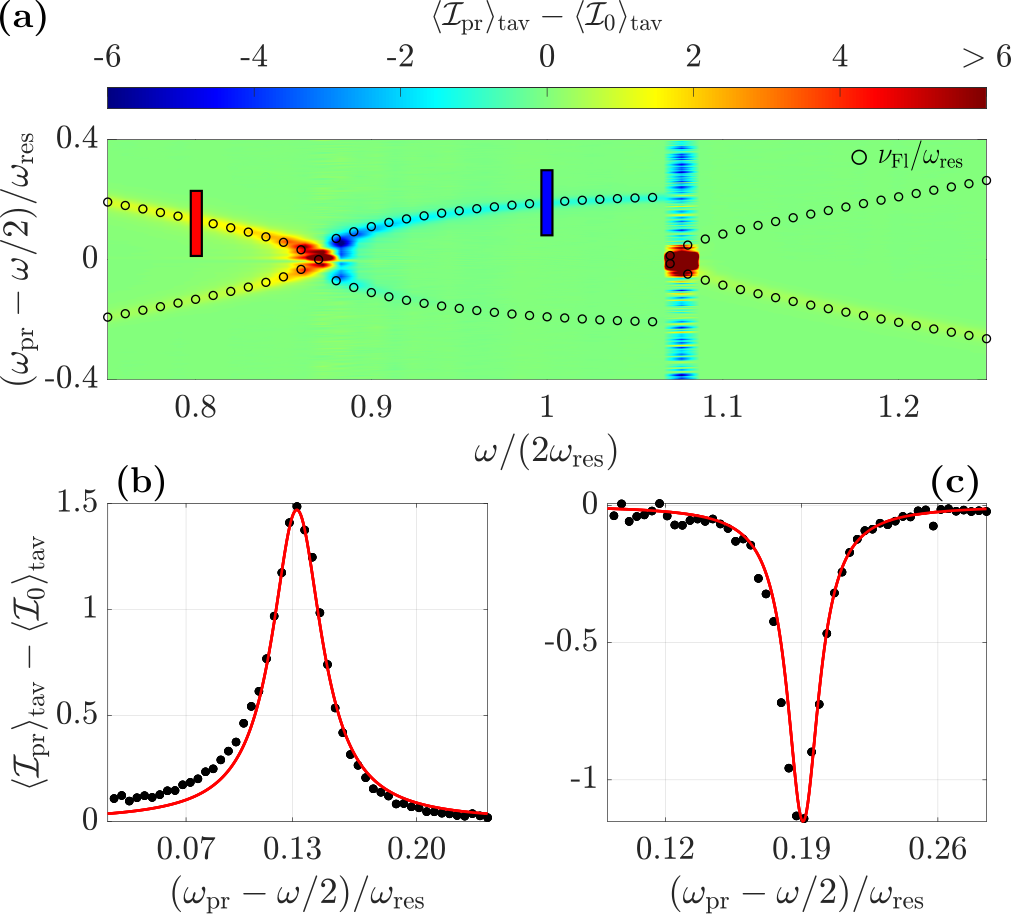}
	\caption{(a) Difference between the time-averaged intensity with probe $\langle\mathcal{I}_\mathrm{pr}\rangle_\mathrm{tav}$ and without $\langle\mathcal{I}_0\rangle_\mathrm{tav}$ obtained from the stochastic simulations (colored background) as a function of the driving frequency $\omega/(2\omega_\text{res})$ and probe driving frequency $(\omega_\text{pr}-\omega/2)/\omega_\text{res}$. The black circles in (a) $(\circ)$ are the oscillation frequencies calculated from Eq.~\eqref{eq8}. (b) Vertical cut in the normal phase at $\omega/(2\omega_\mathrm{res}) = 0.8$ along the upper branch (red rectangle). (c) Vertical cut within the DTC phase at $\omega/(2\omega_\text{res}) = 1$ along the upper branch (blue rectangle). The black dots in (b) and (c) represent data points along the red and blue rectangles, respectively. The red Lorentzians in (b) and (c) are determined from $\gamma_\mathrm{Fl}$ (linewidth) and $\nu_\mathrm{Fl}$ (center frequency). The amplitudes of the red curves are fitted and the other parameters are $\delta_c = \kappa, \, g_1/g_0 = 0.6,\, g_0/g_c = 0.5,$ and $\Delta = 0.1\kappa$.}
	\label{fig:fig3}
\end{figure}

\emph{Probing Excitations.---} 
Now, our goal is to show how one can probe and analyze these excitations in an experiment. The basic idea is to use a weak dynamical probe such that we can extract the excitations by means of linear response theory. For the physical setup we are studying we add the Hamiltonian $\hat{H}_\mathrm{pr} = \eta^*(t)\hat{a} + \hat{a}^\dagger\eta(t)$ to Eq.~\eqref{eq2}. Here, $\eta(t) = \eta_0 e^{-i\omega_\mathrm{pr}t}$, and we set $\eta_0 \in \mathbb{R}$ to be small and tune the probe driving frequency, $\omega_\mathrm{pr}$. To probe excitations around the DTC, we choose $\omega_\mathrm{pr} = \omega/2 + [-0.4,0.4]\omega_\mathrm{res}$, which means that for each value of $\omega$, we adjust the probing frequency to be centered around the oscillation frequency of the DTC. Examples of how this driving affects the DTC are visible in Fig.~\ref{fig:fig1}(c) and (d), as red lines. We simulate the probing scheme using semiclassical truncated Wigner simulations (see SM~\cite{SM} for more details). For each parameter point, we extract the time-averaged photon number with probe, $\langle\mathcal{I}_\mathrm{pr}\rangle_\mathrm{tav} = \int_{0}^{2T} dt\,\langle\hat{a}^\dagger(t)\hat{a}(t)\rangle_\mathrm{pr}/(2T)$, and without probe, $\langle\mathcal{I}_0\rangle_\mathrm{tav} = \int_{0}^{2T} dt\,\langle\hat{a}^\dagger(t)\hat{a}(t)\rangle_{0}/(2T)$, which, in an experiment, can be measured from the transmission. Note that $\langle\mathcal{I}_0\rangle_\mathrm{tav}$ is the Raman-scattered light, which is close to zero in the NP and non-zero in the DTC phase. We then derive the net change of the photon number, $\langle\mathcal{I}_\mathrm{pr}\rangle_\mathrm{tav} - \langle\mathcal{I}_0\rangle_\mathrm{tav}$, to measure the number of excitations created by the probe.

The result of this simulation is visible in Fig.~\ref{fig:fig3}(a) (colored background) as a function of the driving frequency $\omega$ and probing frequency $\omega_{\mathrm{pr}}$. We find narrow spikes (red) in the photon-number change in the NP that represent the amount of injected photons by the drive in the presence of vanishing Raman-scattered light. In addition, we find narrow dips (blue) in the DTC phase, highlighting that the dynamical probing manipulates the superradiant order in a destructive way. As depicted in Fig.~\ref{fig:fig3}(a), we overlay the oscillation frequencies $\nu_\mathrm{Fl}/\omega_\mathrm{res}$ (black circles), calculated from the atom-only stability analysis, and see excellent agreement with the probe-induced responses in both phases. This result first highlights that the Floquet excitation frequencies calculated with atom-only theory provide an accurate description. Second, and more broadly, it shows how the Floquet frequencies $\nu_{\mathrm{Fl}}$ have to be interpreted: they are the probe excitations of the system around the DTC response frequency $\omega/2$.  

We gain insight into the shape of the excitation spectrum in NP (yellow / red regions in Fig.~\ref{fig:fig3}(a)). Since the system predominantly occupies $\ket{\downarrow}$, we can derive a simple Hamiltonian that can be diagonalized and find its eigenenergies (see SM~\cite{SM}). By neglecting dissipation and time-dependent terms, we were able to recover the general behavior of the oscillation frequencies in the NP. Interestingly, this method yields a spectrum that corresponds to the top-left and bottom-right branches, and therefore explains the asymmetry of the excitation spectrum in the NP. The less visible branches can be attributed to the fact that the system can be excited by being driven at $\omega/2$ or $-\omega/2$, where, in Fig.~\ref{fig:fig3}(a), we chose the former. This means that the brighter branches correspond to an excitation directly caused by the probe photon, while the less bright branches stem from a higher-order photon process of absorbing an additional photon from the drive.

We now discuss the features in the DTC phase where the average intensity difference changes from positive to negative. Here, the weak probe destabilizes the DTC, causing it to melt. An explanation for the asymmetric response in the DTC phase is difficult, since we cannot apply any high-frequency approximations because the drive frequency is on the same order as the characteristic energies of the system. Also, the mean-field quantities themselves are time-dependent, so an effective time-independent description is not valid. Near $\omega \approx 1.08/(2\omega_\mathrm{res})$, we see heightened fluctuations near (ii) the discontinuous transition highlighting the bistable character of that parameter regime, as well as a strong response due to the probe being quasi-resonant with $\omega/2$. 

We now study the lineshape of the excitations visible in Figs.~\ref{fig:fig3}(b)–(c) in detail. Figure~\ref{fig:fig3}(b) is a vertical cut of the top branch [red rectangle in Fig.~\ref{fig:fig3}(a)] in the NP for a small range of $\omega_\mathrm{pr}$ at $\omega = 0.8/(2\omega_\mathrm{res})$. The black dots come from the full truncated Wigner simulations, and the red Lorentzian is generated from the atom-only Floquet description. The center of the Lorentzian is at $\nu_\mathrm{Fl}/\omega_\mathrm{res}$, with linewidth $\gamma_\mathrm{Fl}/\omega_\mathrm{res}$. We find excellent agreement, meaning the Floquet theory is able to predict the frequency and linewidth of the excitations. In Fig.~\ref{fig:fig3}(c), the system is in the DTC phase, and probing on resonance causes a decrease in overall intensity. Also, in this phase, we compare the simulated line shape (black dots) with the calculated Lorentzian from Floquet theory (red line), and find excellent agreement. This analysis shows that the developed theory is able to quantitatively describe the line shapes that can be measured in an experiment.

\emph{Conclusion and outlook.---}
In conclusion, we have developed and employed Floquet theory to describe the excitation spectrum of the time-periodic dissipative Dicke model. With this description, we have derived the frequencies and linewidths of the low-frequency collective excitations around the DTC phase. Working with this model has allowed us to analyze the behavior of the excitations across continuous and discontinuous transitions, where we have found fundamentally different behaviors. We then provided tools on how these excitations can be probed in an experiment and how Floquet theory can be used to analyze the measured features. Our work provides an important step towards the analysis of dissipative, time-crystalline phases and dynamical, out-of-equilibrium phases more generally.

We believe that our work provides a solid basis for studying the dynamics of more exotic dynamical phases. A natural next step would be to analyze the excitation spectrum of continuous time crystals that break continuous time-translational symmetries~\cite{Kongkhambut:2022}. Here, we expect the emergence of a gapless, critical mode in the continuous time-crystalline phase. For their description, we believe it will be crucial to include quantum fluctuations as gapless modes, which are usually prone to any type of noise. Further steps could be to analyze the phase with several frequency components, like the recently studied torus bifurcations~\cite{Cosme:2025}, which would require more involved Floquet descriptions. Eventually, systems that spontaneously break symmetries besides time-translational ones, but also continuous spatial symmetries, like the Faraday pattern in Bose–Einstein condensates~\cite{Staliunas:2002,Engels:2007,Liebster:2025}, should be analyzed. Here, it will be interesting to see if new phonon branches emerge, how they can be modified by the driving, and how they can be probed. In conclusion, we believe that our Letter will be a starting point of a series of investigations of low-frequency excitations of dynamical phases.

\begin{acknowledgments}
	\emph{Acknowledgements.---}
	The authors thank T. Schmit for support and discussions and S. Roy, S. Saju, T. Enss, S. Eggert, I. Schneider, J.-M. Giesen, H. Ke\ss{}ler, J. Skulte,  J. Cosme, and C. Kollath for stimulating discussions. This work was funded by the Deutsche Forschungsgemeinschaft (DFG, German Research Foundation) – Project-ID 429529648 – TRR 306 QuCoLiMa (“Quantum Cooperativity of Light and Matter", subproject D02), the DFG Forschergruppe WEAVE ”Quantum many-body
	dynamics of matter and light in cavity QED” - DFG project ID 525057097, the QuantERA project ”QNet: Quantum transport,
	metastability, and neuromorphic applications in Quantum Networks” - DFG project ID 532771420.  SBJ acknowledges support from the Deutsche Forschungsgemeinschaft (DFG, German Research Foundation) under project number 277625399 - TRR 185 (B4) and under Germany’s Excellence Strategy – Cluster of Excellence Matter and Light for Quantum Computing (ML4Q) EXC 2004/1 – 390534769.
\end{acknowledgments}


	\newpage 
	
	\setcounter{equation}{0}
	\setcounter{figure}{0}
	\renewcommand*{\citenumfont}[1]{S#1}
	\renewcommand{\thesection}{S~\arabic{section}}
	\renewcommand{\thesubsection}{\thesection.\arabic{subsection}}
	\makeatletter 
	\def\tagform@#1{\maketag@@@{(S\ignorespaces#1\unskip\@@italiccorr)}}
	\makeatother
	\makeatletter
	\makeatletter \renewcommand{\fnum@figure}
	{\figurename~S\thefigure}
	\onecolumngrid
	\newpage
	\begin{center}
		\textbf{\large Supplemental Material: Collective Excitations of Dissipative Time Crystals}
	\end{center}
	\vspace{1cm}
	\twocolumngrid
	\tableofcontents
\section{Stochastic simulation of the dissipative Dicke model with probe}
In this section we derive the stochastic semiclassical differential equations that are numerically integrated and whose results are presented in Fig.~3 of the main text. We start with the Heisenberg-Langevin equations for the collective spin operators and cavity field
\begin{equation}\label{S1}
	\begin{aligned}
		\frac{d\hat{a}}{dt} &= -i\frac{2g(t)}{\sqrt{N}}\hat{J}^x -i\delta_c\hat{a} - \kappa\hat{a} \\ &- i\eta_0e^{-i(\omega_\mathrm{pr}t - \phi)} + \sqrt{2\kappa}\hat{a}_\mathrm{in}(t),
	\end{aligned}
\end{equation}
\begin{equation}\label{S2}
	\frac{d\hat{J}^x}{dt} = -\Delta\hat{J}^y,
\end{equation}
\begin{equation}\label{S3}
	\frac{d\hat{J}^y}{dt} = \Delta\hat{J}^x - \frac{2g(t)}{\sqrt{N}}(\hat{a} + \hat{a}^\dagger)\hat{J}^z,
\end{equation}
\begin{equation}\label{S4}
	\frac{d\hat{J}^z}{dt} = \frac{2g(t)}{\sqrt{N}}(\hat{a} + \hat{a}^\dagger)\hat{J}^y.
\end{equation}
Where we have introduced the noise operator $\hat{a}_\mathrm{in}$ with zero mean value $\langle\hat{a}_\mathrm{in}\rangle = 0$ and non-vanishing delta-correlated second-order moment $\langle\hat{a}_\mathrm{in}(t)\hat{a}^\dagger_\mathrm{in}(t')\rangle = \delta(t - t')$. The probe $\eta(t) = \eta_0e^{-i(\omega_\text{pr}t - \phi)}$ has strength $\eta_0$ with probe frequency $\omega_\mathrm{pr}$ and phase shift $\phi$. We can express the complex cavity field as Hermitian quadratures $\hat{a}_x = \hat{a}^\dagger + \hat{a}$ and $\hat{a}_p = i(\hat{a}^\dagger - \hat{a})$. This also transforms the noise operator into $\hat{\mathcal{N}}^x = [\hat{a}_\mathrm{in}(t) + \hat{a}_\mathrm{in}^\dagger(t)]$ and $\hat{\mathcal{N}}^p = -i[\hat{a}_\mathrm{in}(t) - \hat{a}^\dagger_\mathrm{in}(t)]$, which fulfills $\langle\hat{\mathcal{N}^\alpha}\rangle = 0$ and $\langle\{\hat{\mathcal{N}}^\alpha(t),\hat{\mathcal{N}}^\beta(t')\}\rangle/2 = \delta_{\alpha\beta}\delta(t - t')$, where $\{.,.\}$ is the anticommutator. The coupled equations of motion are then
\begin{equation}\label{S5}
	\begin{aligned}
		\frac{d\hat{a}_x}{dt} = -\kappa\hat{a}_x + \delta_c\hat{a}_p - 2\eta_0\text{sin}(\omega_\mathrm{pr}t - \phi) + \sqrt{2\kappa}\hat{\mathcal{N}^x}(t),
	\end{aligned}
\end{equation}
\begin{equation}\label{S6}
	\begin{aligned}
		\frac{d\hat{a}_p}{dt} &= -\kappa\hat{a}_p - \delta_c\hat{a}_x - \frac{4g(t)}{\sqrt{N}}\hat{J}^x - 2\eta_0\text{cos}(\omega_\mathrm{pr}t - \phi)\\&+ \sqrt{2\kappa}\hat{\mathcal{N}^p}(t),
	\end{aligned}
\end{equation}
\begin{equation}\label{S7}
	\begin{aligned}
		\frac{d\hat{J}^x}{dt} = -\Delta\hat{J}^y,
	\end{aligned}
\end{equation}
\begin{equation}\label{S8}
	\begin{aligned}
		\frac{d\hat{J}^y}{dt} = \Delta\hat{J}^x - \frac{2g(t)}{\sqrt{N}}\hat{a}_x\hat{J}^z,
	\end{aligned}
\end{equation}
\begin{equation}\label{S9}
	\begin{aligned}
		\frac{d\hat{J}^z}{dt} = \frac{2g(t)}{\sqrt{N}}\hat{a}_x\hat{J}^y.
	\end{aligned}
\end{equation}

The stochastic equations presented here are operator valued, but we seek to exchange these operators with real functions $(\hat{J}^\alpha \rightarrow J^\alpha, \hat{a}_\beta \rightarrow a_\beta)$ using symmetric ordering. In addition, we substitute the quantum noise with classical noise. The stochastic equations simulated in the text are 
\begin{equation}\label{S10}
	\begin{aligned}
		\frac{d{a}_x}{dt} = -\kappa{a}_x + \delta_c{a}_p - 2\eta_0\text{sin}(\omega_\mathrm{pr}t - \phi) + \sqrt{2\kappa}{\mathcal{N}^x}(t),\\
	\end{aligned}
\end{equation}
\begin{equation}\label{S11}
	\begin{aligned}
		\frac{d{a}_p}{dt} &= -\kappa{a}_p - \delta_c{a}_x - \frac{4g(t)}{\sqrt{N}}{J}^x - 2\eta_0\text{cos}(\omega_\mathrm{pr}t - \phi)\\& + \sqrt{2\kappa}{\mathcal{N}^p}(t),\\
	\end{aligned}
\end{equation}
\begin{equation}\label{S12}
	\begin{aligned}
		\frac{d{J}^x}{dt} = -\Delta{J}^y,\\
	\end{aligned}
\end{equation}
\begin{equation}\label{S13}
	\begin{aligned}
		\frac{d{J}^y}{dt} = \Delta{J}^x - \frac{2g(t)}{\sqrt{N}}{a}_x{J}^z,\\
	\end{aligned}
\end{equation}
\begin{equation}\label{S14}
	\begin{aligned}
		\frac{d{J}^z}{dt} = \frac{2g(t)}{\sqrt{N}}{a}_x{J}^y.
	\end{aligned}
\end{equation}
For the results shown in this paper we consider the initial state with $\langle J^x\rangle =0= \langle J^y\rangle$, $\langle J^z\rangle = -N/2$, with the cavity in vacuum $\langle a_x\rangle = \langle a_p\rangle = 0$. To model quantum fluctuations in the stochastic variables, we initialize them by independent Gaussian random variables with $\langle a_x^2\rangle = \langle a_p^2\rangle = 1$, $\langle (J^x)^2 \rangle = \langle (J^y)^2 \rangle = N/4$, and $\langle (J^z)^2 \rangle = \langle J^z \rangle^2 = N^2/4$. Using the these initial conditions we numerically integrate the stochastic differential equations and average them over many initialization. 
\section{Phase transition boundaries}
\begin{figure}
	\centering
	\includegraphics[width=1\columnwidth]{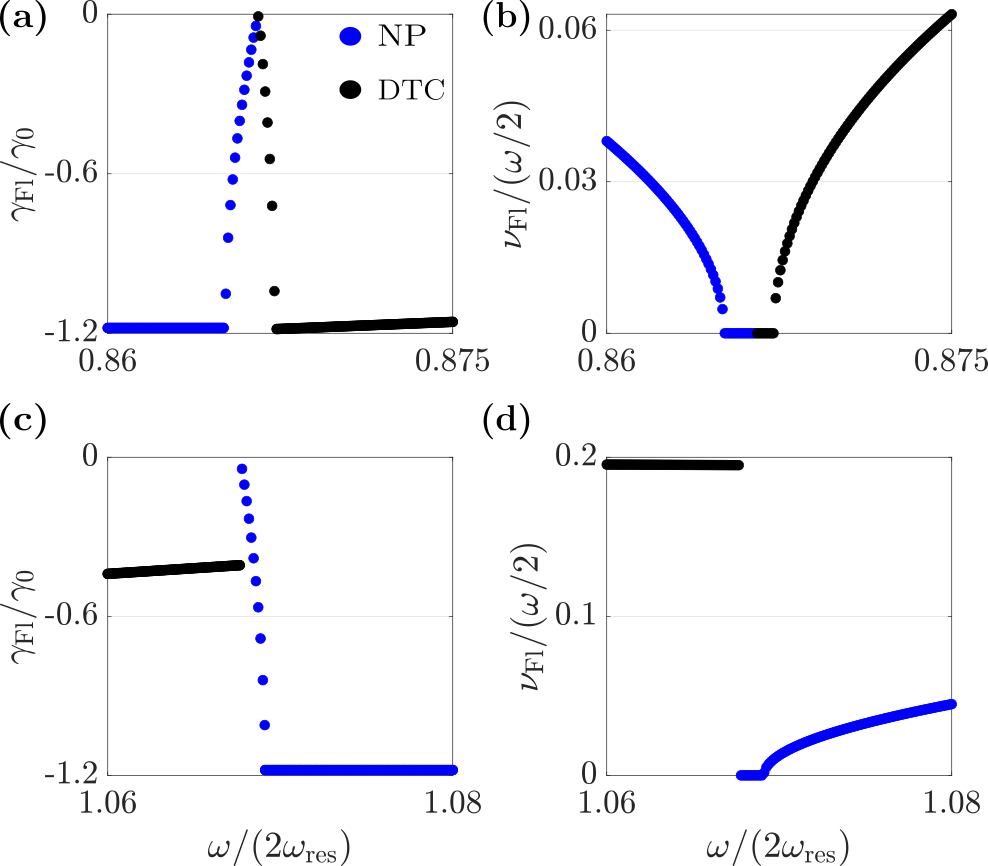}
	\caption{(a)-(b) Enhanced simulations around the continuous transition and (c)-(d) around the discontinuous transition. The discrete dissipative time crystal phase (DTC) is indicated by the black points and the normal phase (NP) is indicated by the blue points. The other parameters are the same as in Fig.~2 of the main text.}
	\label{fig:figzoomed}
\end{figure}

In this section we show enhanced results around the continuous and discontinuous transitions that are present in Fig.~2(c) and (d) of the main text. In Fig.\ref{fig:figzoomed}(a) and (b) we highlight the continuous nature of the transition by seeing the relaxation rate $\gamma_\text{Fl}$ going to zero and the oscillation frequency becoming gapless. While in Fig.\ref{fig:figzoomed}(c) and (d) we see the discontinuous nature where the DTC phase (black points) jumps to the normal phase (blue points). 

\section{Elimination of the cavity field with probe}
In this section we provide an explicit calculation of the elimination of the cavity degrees of freedom. This is done by performing a Schrieffer-Wolf transformation by going to the displaced frame defined by $\hat{D}(t) = \text{exp}(\hat{a}^\dagger\hat{\beta}(t) - \hat{\beta}^\dagger(t)\hat{a})$ ~\cite{Jaeger:2022}. We focus on the derivation of the transformation operator $\hat{\beta}(t)$, whose evolution is determined by 
\begin{equation}\label{S15}
	\begin{aligned}
		i\frac{\partial\hat{\beta}}{\partial t} = (\delta_c - i\kappa)\hat{\beta} + \frac{g(t)}{\sqrt{N}}(\hat{J}^+ + \hat{J}^-) + [\Delta\hat{J}^z, \hat{\beta}] + \eta(t)\hat{\mathbb{I}},
	\end{aligned}
\end{equation}
where $\hat{\mathbb{I}}$ is the identity operator.
To solve this equation we make the ansatz $\hat{\beta}(t) = c_+(t)\hat{J}^+ + c_-(t)\hat{J}^- + c_\mathrm{pr}(t)\hat{\mathbb{I}}$, and need to solve for the time-dependent coefficients. Their equations of motion are 
\begin{equation}\label{S16}
	\begin{aligned}
		i\frac{dc_\pm}{dt} = (\delta_c \pm \Delta - i\kappa)c_\pm + \frac{g(t)}{\sqrt{N}},
	\end{aligned}
\end{equation}
\begin{equation}\label{S17}
	\begin{aligned}
		i\frac{dc_\mathrm{pr}}{dt} = (\delta_c -i\kappa)c_\mathrm{pr} + \eta(t)\hat{\mathbb{I}},
	\end{aligned}
\end{equation}
which can be formally solved by
\begin{equation}\label{S18}
	\begin{aligned}
		c_\pm = \frac{-i}{\sqrt{N}}\int^t_0 e^{[-i(\delta_c \pm \Delta) - \kappa]\tau}g(t - \tau)d\tau,
	\end{aligned}
\end{equation}
and
\begin{equation}
	\begin{aligned}
		c_\mathrm{pr} = -i\int_0^te^{[-i\delta_c - \kappa]\tau}\eta(t - \tau)d\tau.
	\end{aligned}
\end{equation}
We drop the homogeneous solution since it can be discarded after times $t \gg \kappa^{-1}$. Assuming $\omega \ll \kappa, \delta_c$ we can use that $g(t)$ changes slowly so that we can approximate it as $g(t - \tau) \approx g(t) - \tau\dot{g}(t)$ and get 
\begin{equation}\label{S19}
	\begin{aligned}
		c_\pm &\approx -\frac{g(t)}{\sqrt{N}[\delta_c \pm \Delta - i\kappa]} - \frac{i\dot{g}(t)}{\sqrt{N}[\delta_c \pm \Delta - i\kappa]^2},\\
		&\approx -\frac{1}{\sqrt{N}}\Bigg(\frac{g(t)}{\delta_c - \kappa} + \frac{i\dot{g}(t)}{[\delta_c - i\kappa]^2} \mp \frac{\Delta g(t)}{[\delta_c - i\kappa]^2}\Bigg),
	\end{aligned}
\end{equation}
where we used $\kappa t \gg 1$ for the first equation and $\Delta, \omega \ll \delta_c, \kappa$ in the second equation. The probe coefficient is approximated to  
\begin{equation}\label{S20}
	\begin{aligned}
		c_\mathrm{pr} \approx -\frac{\eta(t)}{\delta_c - i\kappa} - \frac{i\dot{\eta}(t)}{[\delta_c - i\kappa]^2},
	\end{aligned}
\end{equation}
and we used $\omega_\mathrm{pr} \ll \kappa, \delta_c$ allowing for the expansion $\eta(t - \tau) \approx \eta(t) - \tau\dot{\eta}(t)$.

\section{Derivation of the mean-field equations with probe}
In this section we derive the mean-field equations with the addition of a probe. Starting from the master equation shown in the main text (Eq.~1) and utilizing the elimination of the photonic degrees, we arrive at the effective atom-only master equation $(\hbar = 1)$
\begin{equation}\label{S21}
	\begin{aligned}
		\frac{\partial \hat{\rho}_\mathrm{at}}{\partial t} = -i[\hat{H}_\mathrm{at}, \hat{\rho}_\mathrm{at}] - \kappa\Big(\hat{\beta}^\dagger\hat{\beta}\hat{\rho}_\mathrm{at} + \hat{\rho}_\mathrm{at}\hat{\beta}^\dagger\hat{\beta} - 2\hat{\beta}\hat{\rho}_\mathrm{at}\hat{\beta}^\dagger\Big).
	\end{aligned}
\end{equation}
The coherent time evolution of the atoms is governed by the effective atom-only Hamiltonian
\begin{equation}\label{S22}
	\begin{aligned}
		\hat{H}_\mathrm{at} = \Delta\hat{J}^z + \frac{g(t)}{2\sqrt{N}}\Big(\hat{\beta}^\dagger[\hat{J}^+ + \hat{J}^- + \eta(t)\hat{\mathbb{I}}] + \text{H.c.}\Big).
	\end{aligned}
\end{equation}
We use a Schwinger mapping and introduce bosonic operators $\hat{\varphi}_\downarrow$ and $\hat{\varphi}_\uparrow$ such that 
\begin{align}
	\hat{J}^+=&\hat{\varphi}_\uparrow^\dag\hat{\varphi}_{\downarrow}\\
	\hat{J}^z=&\frac{\hat{\varphi}_\uparrow^\dag\hat{\varphi}_{\uparrow}-\hat{\varphi}_\downarrow^\dag\hat{\varphi}_{\downarrow}}{2}.
\end{align}

The mean-field equations of motion for the two internal states are ${\varphi}_{s} = \langle\hat{\varphi}_{s}\rangle = \text{Tr}(\hat{\varphi}_s\hat{\rho}_\mathrm{at})$ for $s = \{\,\downarrow, \uparrow\}$, then using the cyclic property of the trace we obtain
\begin{equation}\label{S23}
	\begin{aligned}
		\frac{d\langle\hat{\varphi}_s\rangle}{dt} = -i\langle[\hat{\varphi}_s, \hat{H}_\mathrm{at}]\rangle - \kappa\langle\hat{\beta}^\dagger[\hat{\beta},\hat{\varphi}_s] + [\hat{\varphi}_s,\hat{\beta}^\dagger]\hat{\beta}\rangle.
	\end{aligned}
\end{equation}
Now by substituting the ansatz $\hat{\beta}(t) = c_+(t)\hat{J}^+ + c_-(t)\hat{J}^- + c_\mathrm{pr}(t)\hat{\mathbb{I}}$ and factorizing higher order moments in the bosonic operators we find 
\begin{equation}\label{S24}
	\begin{aligned}
		\frac{d\varphi_\downarrow}{dt} = &-i\frac{g(t)}{\sqrt{N}}\text{Re}\big(c_+ + c_-\big)\abs{\varphi_\uparrow}^2\varphi_\downarrow\\ &-i\frac{g(t)}{\sqrt{N}}\big(c^*_+ + c_-\big)(\varphi_\uparrow)^2\varphi_\downarrow^* \\ &-i\Big(\frac{g(t)}{\sqrt{N}}\text{Re}(c_\mathrm{pr}) + \frac{\eta(t)}{2}c^*_+ + \frac{\eta^*(t)}{2}c_-\Big)\varphi_\uparrow \\ &+ \kappa\big(\abs{c_-}^2 - \abs{c_+}^2\big)\abs{\varphi_\uparrow}^2\varphi_\downarrow \\ &+ \kappa\big(c_\mathrm{pr}^*c_- - c_\mathrm{pr}c_+^*\big)\varphi_\uparrow,
	\end{aligned}
\end{equation}
\begin{equation}\label{S25}
	\begin{aligned}
		\frac{d\varphi_\uparrow}{dt} = &-i\Delta\varphi_\uparrow -i\frac{g(t)}{\sqrt{N}}\text{Re}\big(c_+ + c_-\big)\abs{\varphi_\downarrow}^2\varphi_\uparrow\\ &-i\frac{g(t)}{\sqrt{N}}\big(c_+ + c_-^* \big)(\varphi_\downarrow)^2\varphi_\uparrow^* \\ &-i\Big(\frac{g(t)}{\sqrt{N}}\text{Re}(c_\mathrm{pr}) + \frac{\eta(t)}{2}c^*_- + \frac{\eta^*(t)}{2}c_+\Big)\varphi_\downarrow \\ &+ \kappa\big(\abs{c_+}^2 - \abs{c_-}^2\big)\abs{\varphi_\downarrow}^2\varphi_\uparrow \\ &+ \kappa\big(c_\mathrm{pr}^*c_+ - c_\mathrm{pr}c_-^*\big)\varphi_\downarrow.
	\end{aligned}
\end{equation}
Using  Eqs.\eqref{S19} and \eqref{S20} we can introduce
\begin{equation}\label{S30}
	\begin{aligned}
		V_0 = \frac{2\delta_c g^2(t)}{\delta_c^2 + \kappa^2} - \frac{4\delta_c \kappa g(t)\dot{g}(t)}{[\delta_c^2 + \kappa^2]^2},
	\end{aligned}
\end{equation}
\begin{equation}\label{S31}
	\begin{aligned}
		V_1 = \frac{4\delta_c\Delta\kappa g^2(t)}{[\delta_c^2 + \kappa^2]^2},
	\end{aligned}
\end{equation}
\begin{equation}\label{S32}
	\begin{aligned}
		V_2 = \frac{2g(t)\eta_0}{\sqrt{N}[\delta^2_c + \kappa^2]}\Big(\delta_c\text{cos}(\Theta) + \kappa\text{sin}(\Theta)\Big),
	\end{aligned}
\end{equation}
where $\Theta = \omega_{pr}t - \phi$ to get
\begin{equation}\label{S33}
	\begin{aligned}
		\frac{d\varphi_\downarrow}{dt} = i\frac{V_0 - iV_1}{N}\abs{\varphi_\uparrow}^2\varphi_\downarrow + i\frac{V_0 + iV_1}{N}\varphi^2_\uparrow \varphi^*_\downarrow + i{V_2}\varphi_\uparrow,
	\end{aligned}
\end{equation}
\begin{equation}\label{S34}
	\begin{aligned}
		\frac{d\varphi_\uparrow}{dt} = &-i\bigg(\Delta - \frac{V_0 + iV_1}{N}\abs{\varphi_\downarrow}^2\bigg)\varphi_\uparrow\\ &+ i\frac{V_0 - iV_1}{N}\varphi^2_\downarrow\varphi^*_\uparrow + i{V_2}\varphi_\downarrow.
	\end{aligned}
\end{equation}
It is important to note that we have treated the probe as a perturbation.

\begin{figure}
	\centering
	\includegraphics[width=1\columnwidth]{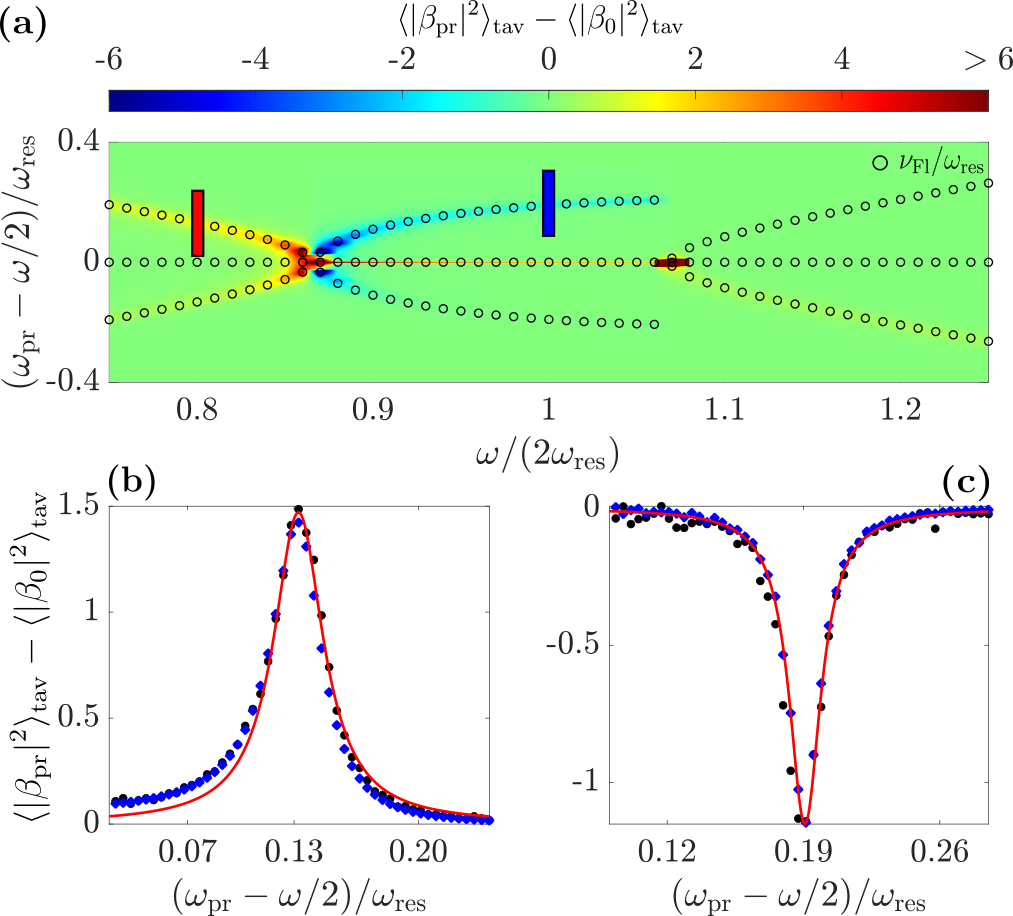}
	\caption{(a) Difference between the averaged intensity with probe $\langle|{\beta}_\mathrm{pr}|^2\rangle_\mathrm{avg}$ and the averaged intensity without probe $\langle|{\beta}_0|^2\rangle_\mathrm{avg}$ obtained from the atom-only mean-field simulations (colored background) as a function of the driving frequency $\omega/(2\omega_\text{res})$ and probe driving frequency $(\omega_\text{pr}-\omega/2)/\omega_\text{res}$. The black circles in (a) $(\circ)$ are the oscillation frequencies calculated from the atom-only stability analysis. (b) Vertical slice in the normal phase at $\omega/(2\omega_\mathrm{res}) = 0.8$ along the upper branch (red rectangle). (c) Vertical slice within the DTC phase at $\omega/(2\omega_\text{res}) = 1$ along the upper branch (blue rectangle). The black dots in (b) and (c) represent data from the stochastic simulations. The blue points in (b) and (c) is data from atom-only mean-field simulations. The red Lorentzians in (b) and (c) determined from $\gamma_\mathrm{Fl}$ (linewidth) and $\nu_\mathrm{Fl}$ (center). The amplitudes of the red curves were fitted and the other parameters are $\delta_c = \kappa, \, g_1/g_0 = 0.6,\, g_0/g_c = 0.5,$ and $\Delta = 0.1\kappa$.}
	\label{fig:figMFprobe}
\end{figure}

In order to get the three coupled equations shown in the main text we define the collective mean-field spin operators as 
\begin{equation}\label{S35}
	\begin{aligned}
		2J^x = \varphi_\uparrow^*\varphi_\downarrow + \varphi_\downarrow^*\varphi_\uparrow,
	\end{aligned}
\end{equation}
\begin{equation}\label{S36}
	\begin{aligned}
		2J^y = i(\varphi_\downarrow^*\varphi_\uparrow - \varphi_\uparrow^*\varphi_\downarrow),
	\end{aligned}
\end{equation}
\begin{equation}\label{S37}
	\begin{aligned}
		2J^z = \varphi_\uparrow^*\varphi_\uparrow - \varphi_\downarrow^*\varphi_\downarrow.
	\end{aligned}
\end{equation}
By taking the time derivative of Eqs.~\eqref{S35}-\eqref{S37} and substituting Eqs.~\eqref{S33} and \eqref{S34} we get
\begin{equation}\label{S38}
	\begin{aligned}
		\frac{dJ^x}{dt} = -\Delta J^y,
	\end{aligned}
\end{equation}
\begin{equation}\label{S39}
	\begin{aligned}
		\frac{dJ^y}{dt} = \Delta J^x + \frac{4V_0J^xJ^z}{N} + \frac{4V_1J^yJ^z}{N} + {2{V_2} J^z},
	\end{aligned}
\end{equation}
\begin{equation}\label{S40}
	\begin{aligned}
		\frac{dJ^z}{dt} = -\frac{4V_0J^xJ^y}{N} - \frac{4V_1J^yJ^y}{N} - {2{V_2}J^y}.
	\end{aligned}
\end{equation}
Discarding terms containing ${V}_2$ we get the mean-field description shown in the main text. Using Eqs.\eqref{S38}-~\eqref{S40} in Fig.~\ref{fig:figMFprobe} we replicate the results shown in Fig.3 of the main text. In Fig.~\ref{fig:figMFprobe}(a) we see very similar asymmetries within the response branches but another response on resonance at $\omega_\text{pr} = \omega/2$ within the DTC phase. In Fig.~\ref{fig:figMFprobe}(b)-(c) we see great agreement between the stochastic simulation results (black dots), the red Lorentzian from the atom-only stability analysis, and the atom-only mean-field equations with probe (blue dots). This highlights the validity of the atom-only description.  

\section{Derivation of the fluctuation matrix with probe}
In this section we derive the fluctuation matrix shown in the stability analysis section of the main text. We first recast Eqs.\eqref{S38}-\eqref{S40} into 
\begin{align}
	\frac{d\vec{v}}{dt}= {\bf M}[\vec{v}]\vec{v}+\Sigma_{\mathrm{pr}}\vec{v}
\end{align}
with $\vec{v}=(J^x,J^y,J^z)^T$ and
the following matrices
\begin{equation}\label{S41}
	\begin{aligned}
		{\bf M}[\vec{v}]=& \begin{pmatrix}
			0 & -\Delta & 0 \\ \\ \Delta + \frac{4V_0J^z_0}{N} & \frac{4V_1J^z_0}{N} & \frac{4V_0J^x_0}{N} + \frac{4V_1J^y_0}{N} \\ \\ -\frac{4V_0J^y_0}{N} & -\frac{4V_0J^x_0}{N} - \frac{4V_1J^y_0}{N}& 0   
		\end{pmatrix},
	\end{aligned}
\end{equation}
and
\begin{equation}\label{S43}
	\begin{aligned}
		\Sigma_{\mathrm{pr}}=&\begin{pmatrix}
			0 & 0 & 0 \\ \\ 0& 0& 2{V}_2\\ \\ 0& - 2{V}_2& 0   
		\end{pmatrix}.
	\end{aligned}
\end{equation}
Then we use the ansatz of expressing the collective variables as a sum of a large periodic contribution $J^\alpha_0(t) = J^\alpha_0(t + T)$ and a small contribution due to quantum fluctuations $\delta J^\alpha = J^\alpha - J^\alpha_0$. Defining a vector for each contribution $v_0 = (J^x_0, J^y_0, J^z_0)^\text{T}$ and $\delta v = (\delta J^x, \delta J^y, \delta J^z)^\text{T}$ and substituting this ansatz into Eq.\eqref{S41} we get
\begin{equation}\label{S42}
	\begin{aligned}
		\frac{d(\Vec{v}_0 + \delta\Vec{v})}{dt} &= \textbf{M}[\Vec{v}_0 + \delta\Vec{v}](\Vec{v}_0 + \delta\Vec{v})+\Sigma_{\mathrm{pr}}(\Vec{v}_0 + \delta\Vec{v})\\
		&= \textbf{M}[\Vec{v}_0]\Vec{v}_0+\Sigma_{\mathrm{pr}}\Vec{v}_0 + \Sigma[\Vec{v}_0]\delta\Vec{v} + \mathcal{O}(\delta\Vec{v}^2). 
	\end{aligned}
\end{equation}
We are focused on the linear response of the atoms that is how quantum fluctuations of first-order behave around the long time solutions ($J^x_0, J^y_0, J^z_0)$, therefore we neglect terms that are second-order in fluctuations $\mathcal{O}(\delta\Vec{v}^2)$ or higher. We can split the fluctuation matrix $\Sigma[\Vec{v}_0] = \Sigma_0(t) + \Sigma_\text{pr}$ into two parts, one part containing the original mean-field terms $\Sigma_0$ and another that includes the probe terms $\Sigma_\mathrm{pr}$. The corresponding matrix $\Sigma_0$ is defined as
\begin{equation}\label{S44}
	\begin{aligned}
		\Sigma_0 = \begin{pmatrix}
			0 & -\Delta & 0 \\ \\ \Delta + \frac{4V_0J^z_0}{N} & \frac{4V_1J^z_0}{N} & \frac{4V_0J^x_0}{N} + \frac{4V_1J^y_0}{N}\\ \\ -\frac{4V_0J^y_0}{N} & -\frac{4V_0J^x_0}{N} - \frac{8V_1J^y_0}{N} & 0   
		\end{pmatrix}.
	\end{aligned}
\end{equation}
By ignoring the $\Sigma_\mathrm{pr}$ contribution we recover the fluctuation matrix presented in the main text.
\section{Numerical simulation of the fluctuation matrix}
In this section we give additional information on how we calculated the Floquet eigenvalues plotted in the main text. We begin by writing the equations of motion of the quantum fluctuations in a compact manner
\begin{equation}\label{S45}
	\begin{aligned}
		\frac{d\delta\Vec{v}}{dt} = \Sigma(t)\delta\Vec{v},
	\end{aligned}
\end{equation}
where the fluctuation matrix $\Sigma(t)$ is a time-periodic object. In order to find said eigenvalues of $\Sigma$ we need to evolve Eqs.\eqref{S38}-\eqref{S40} (neglecting probe terms) for a sufficiently long time $\tau_0$ such that the system has either reached a steady state or limit cycle depending on if the system is in the normal or DTC phase. Since we are focused on determining the excitation spectrum of the DTC, the collective variable will be $2T$-periodic. Therefore, we discretize the number of points within two periods to be 
$N_\mathrm{cut}$ and take the last $N_\mathrm{cut}$ values for the collective variables and other time-dependent parameters. The evolution of the fluctuation matrix $\Sigma$ can be expressed as a time-evolution operator defined by 
\begin{equation}\label{S46}
	\begin{aligned}
		\Sigma(\tau_0 + 2T, \tau_0) = \mathcal{T}\text{exp}\Big[\int_{\tau_0}^{\tau_0 + 2T}\Sigma(\tau)d\tau\Big],
	\end{aligned}
\end{equation}
where $\mathcal{T}$ denotes time-ordering. The expression above can be computationally costly so we perform a Trotter decomposition  by dividing the timespan $[\tau_0, \tau_0 + 2T]$ into $N_\mathrm{cut}$ segments $\tau_j = \tau_0 + j\delta\tau$, allowing us to rewrite Eq.\eqref{S46} as 
\begin{equation}\label{S47}
	\begin{aligned}
		\Sigma(\tau_0 + 2T, \tau_0) = \prod_{j = 0}^{N_\mathrm{cut}-1}\text{exp}\Big[\Sigma(\tau_j)\delta\tau\Big],
	\end{aligned}
\end{equation}
where $\delta\tau = 2T/N_\mathrm{cut}$ is the timestep. This method leads to an error of $\mathcal{O}(\delta\tau^2)$ so we choose a sufficiently large $N_\mathrm{cut}$ to minimize the error. The complex Floquet eigenvalues used in the main text are found by diagonalizing Eq.\eqref{S47} which yields eigenvalues of the form $ \mu_i = e^{\lambda_\mathrm{Fl,i}2T}$, whereby taking the natural logarithm we get $\lambda_\text{Fl,i} = \ln{(\mu_i)}/2T = \gamma_\text{Fl,i} - i\nu_\text{Fl,i}$. 
\section{Approximation of the excitation spectrum in the normal phase}
\begin{figure}
	\centering
	\includegraphics[width=1\columnwidth]{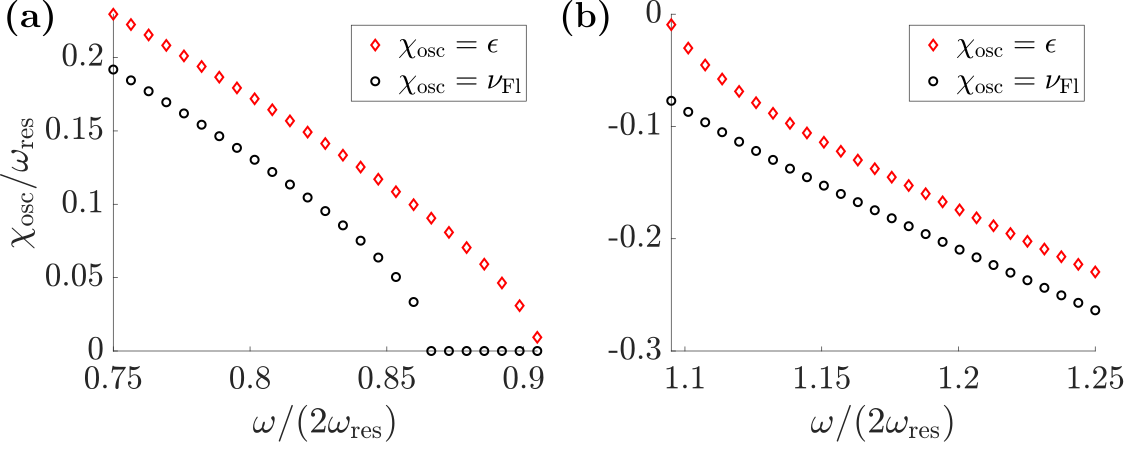}
	\caption{(a)-(b) Comparison of the oscillation frequencies in the normal phase. The oscillation frequencies $\nu_\text{Fl}$ from the fluctuation matrix $\Sigma_0$ Eq.\eqref{S44} are shown as black circles, while the oscillation frequencies $\epsilon$ from Eq.\eqref{S52} are shown as red diamonds and are plotted as a function of the driving frequency $\omega/(2\omega_\mathrm{res})$. The other parameters are the same as in Fig.\ref{fig:figMFprobe}}
	\label{fig:figBogo}
\end{figure}
In this section we provide the derivation and approximations used to find an approximate form of the excitation spectrum in the normal phase. The goal of this calculation is to derive a simple Hamiltonian expression that can be easily diagonalized and compare the resulting oscillation frequency to the oscillation frequency found using Eq.\eqref{S44}. We begin with the effective atom-only Hamiltonian 
\begin{equation}\label{S48}
	\begin{aligned}
		\hat{H}_\mathrm{at} = \Delta\hat{J}^z + \frac{g(t)}{\sqrt{N}}\big(\hat{\beta}^\dagger\hat{J}^x + \hat{J}^x\hat{\beta}\big).
	\end{aligned}
\end{equation}
Within the normal phase the system macroscopically occupies the $\ket{\downarrow}$ state, retardation effects are negligible, and $\Delta \ll \kappa, \delta_c$. Therefore, the time-dependent coefficients $c_+(t)$ and $c_-(t)$ found within the ansatz of $\hat{\beta}(t) = c_+(t)\hat{J}^+ + c_-(t)\hat{J}^-$ are equivalent and only contain their first terms (see Eq.\eqref{S19}). The $\hat{\beta}$ operator then becomes 
\begin{equation}\label{S49}
	\begin{aligned}
		\hat{\beta} = \frac{-2g(t)\delta_c}{\sqrt{N}[\delta_c^2 + \kappa^2]}\hat{J}^x,
	\end{aligned}
\end{equation}
and is equal to its Hermitian conjugate. Upon substituting Eq.\eqref{S49} into Eq.\eqref{S48} and approximating $g^2(t) \approx g_0^2 + 2g_0g_1\text{cos}(\omega t)$ under the assumption that $g_1 \ll g_0$ we get
\begin{equation}\label{S50}
	\begin{aligned}
		\hat{H}_\mathrm{at} = \Delta\hat{J}^z - \Bigg[\frac{4\mathcal{A}}{N} + \frac{4\mathcal{B}}{N}\text{cos}(\omega t)\Bigg](\hat{J}^x)^2,
	\end{aligned}
\end{equation}
where $\mathcal{A} = {g_0^2\delta_c}/{[\delta_c^2 + \kappa^2]}$ and $\mathcal{B} = 2g_0g_1\delta_c/{[\delta_c^2 + \kappa^2]}$.
We now express the collective angular momentum components as bosonic operators $\hat{\varphi}_s \,(\hat{\varphi}^\dagger_s),\, s = \{\downarrow, \uparrow\}$ (boson representation) where we destroy (create) an atom in the $\ket{s}$ state. Being in the normal phase allows us assume $\hat{\varphi}_\uparrow \ll \sqrt{N}$ and $\hat{\varphi}_\downarrow \approx \sqrt{N}$. The new Hamiltonian then reads (ignoring offsets)
\begin{equation}\label{S51}
	\begin{aligned}
		\hat{H}_\mathrm{at} = \Delta\hat{\varphi}_\uparrow^\dagger\hat{\varphi}_\uparrow - \big[\mathcal{A} +  \mathcal{B}\text{cos}(\omega t)\big](\hat{\varphi}^\dagger_\uparrow + \hat{\varphi}_\uparrow)^2.
	\end{aligned}
\end{equation}
We now perform a Bogoliubov transformation by expressing $\hat{\varphi}_\uparrow$ in a new basis of bosonic operators defined by $\hat{\varphi}_\uparrow = \text{cosh}(\Theta_\mathrm{BT})\hat{b}_\uparrow + \text{sinh}(\Theta_\mathrm{BT})\hat{b}^\dagger_\uparrow$ and $\hat{a}_\uparrow^\dagger = \text{cosh}(\Theta_\mathrm{BT})\hat{b}^\dagger_\uparrow + \text{sinh}(\Theta_\mathrm{BT})\hat{b}_\uparrow$ in order to diagonalize the time-independent part $\hat{H}_{\mathrm{TI}}$ of the Hamiltonian which yields
\begin{equation}
	\begin{aligned}
		\hat{H}_{\mathrm{TI}} = \omega_{\mathrm{res}}\hat{b}^\dagger_\uparrow\hat{b}_\uparrow. 
	\end{aligned}
\end{equation}
The angle $\Theta_\mathrm{BT}$ is determined by setting the coefficients in front of the terms $\hat{b}^\dagger_\uparrow\hat{b}^\dagger_\uparrow$ and $\hat{b}\hat{b}$ to zero and reads $\Theta_\mathrm{BT} = \text{coth}^{-1}([\Delta/2\mathcal{A}] - 1)$.
In order to remove the time-dependence in Eq.\eqref{S51} we perform two different frame rotations. The first one is defined by $\hat{U}_1(t) = \text{exp}(-i\omega_\mathrm{res}\hat{b}^\dagger_\uparrow\hat{b}_\uparrow t)$, while the second one is defined by $\hat{U}_2(t) = \text{exp}(-i[-\omega_\mathrm{res} + \omega/2] \hat{b}^\dagger_\uparrow\hat{b}_\uparrow t)$. In addition to these frame rotations we apply a rotating-wave approximation based on $\omega \gg \omega_\mathrm{res} - \omega/2$. This process yields a time-independent Hamiltonian but is still not diagonal, so we apply another Bogoliubov transformation defined by $\hat{b}_\uparrow = \text{cosh}(\phi_\mathrm{BT})\hat{c}_\uparrow + \text{sinh}(\phi_\mathrm{BT})\hat{c}^\dagger_\uparrow$ and $\hat{b}^\dagger_\uparrow = \text{cosh}(\phi_\mathrm{BT})\hat{c}^\dagger_\uparrow + \text{sinh}(\phi_\mathrm{BT})\hat{c}_\uparrow$ which leads us to the desired oscillation frequency  
\begin{equation}\label{S52}
	\begin{aligned}
		\epsilon = &-\mathcal{B}\big[\text{cosh}(2\Theta_\mathrm{BT}) + \text{sinh}(2\Theta_\mathrm{BT})\big]\text{sinh}(2\phi_\mathrm{BT})\\ &+ (\omega_\mathrm{res} - {\omega}/2)\text{cosh}(2\phi_\mathrm{BT}).
	\end{aligned}
\end{equation}
The second Bogoliubov transformation introduces the angle $\phi_\mathrm{BT} = \text{tanh}^{-1}(\mathcal{B}/[\omega_\mathrm{res} - \omega/2])$ and was found using the same logic as for $\Theta_\mathrm{BT}$. We use Eq.\eqref{S52} and compare it to the numerical simulations of Eq.\eqref{S44} as a function of driving frequency $\omega/(2\omega_\mathrm{res})$ in Fig.\ref{fig:figBogo}. In both Fig.\ref{fig:figBogo}(a) and (b) we see the oscillation frequencies from the approximated Hamiltonian $\epsilon$ (red diamonds) are in good qualitative agreement with the oscillation frequencies $\nu_\mathrm{Fl}$ from the stability analysis. By neglecting dissipation, time-dependent terms, and energy offsets we can see the threshold is altered. However, what we find remarkable is that Eq.\eqref{S52} predicts both positive and negative frequencies depending on if parameter regime is before or after the DTC phase. These positive and negative frequencies give us insight as to why we see an asymmetry of the responses in Fig.~3 of the main text. 
\begin{figure}
	\centering
	\includegraphics[width=1\columnwidth]{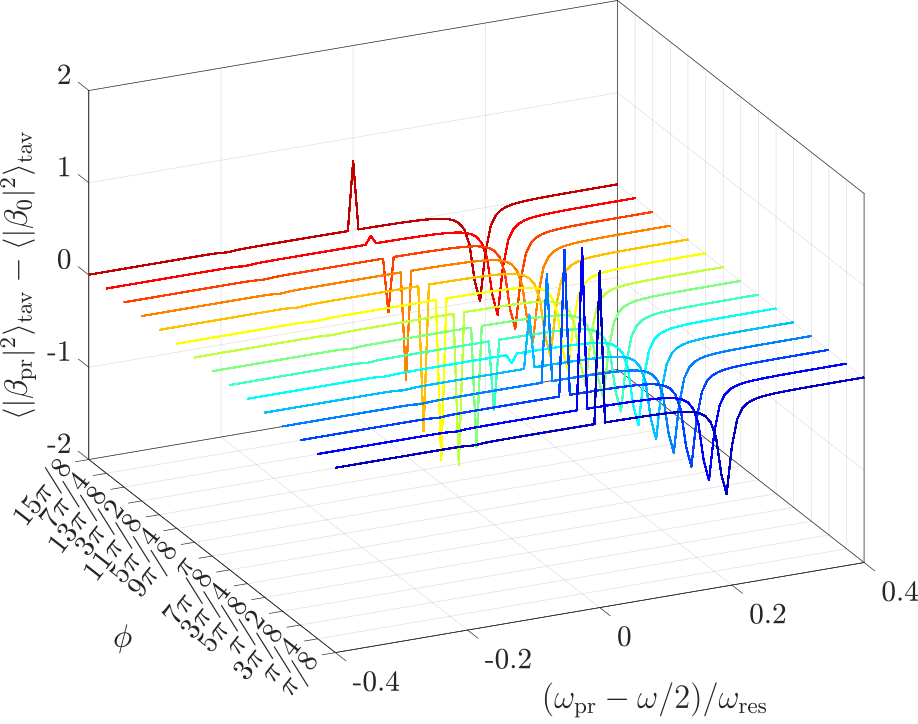}
	\caption{Atom-only mean-field with probe simulation results for the difference between averaged intensity with probe and averaged intensity without probe as a function of probe driving frequency $(\omega_\mathrm{pr} - \omega/2)\omega_\mathrm{res}$ and probe phase $\phi$. The parameters are $\omega = 2\omega_\mathrm{res}$ and the rest are the same as in Fig.\ref{fig:figMFprobe} }
	\label{fig:figMFphase}
\end{figure}
\section{Impact of probe phase on DTC phase}
In this section we discuss the effect of the probe phase $\phi$ on the DTC. In Fig.~\ref{fig:figMFphase} we plot the difference between the averaged intensity with probe and without probe as a function of the probe frequency and probe phase. We focus on a full vertical slice at $\omega = 2\omega_\mathrm{res}$ of Fig.~\ref{fig:figMFprobe}. We see the excitations induced by the probe at $\approx \pm 0.2\omega_\mathrm{res}$ are unaffected by the probe phase. In contrast, when the probe drive is resonant with $\omega/2$, we see that the probe phase has an impact. By changing the probe phase we can gain insight as to what the phase of the light from the external drive being scattered is. If we see an increase in the intensity this means we are phase matching the scattered light resulting in constructive interference. For a difference in intensities that is maximally negative the probe phase will be $\pi$-phase shifted from the phase of the scattered light resulting in destructive interference.    
\bibliography{references.bib}
\end{document}